\PassOptionsToPackage{table,xcdraw}{xcolor}

\documentclass[screen,authorversion]{acmart}

\AtBeginDocument{%
  }

\copyrightyear{2024}
\acmYear{2024}
\setcopyright{rightsretained}
\acmConference[LAK '24]{The 14th Learning Analytics and Knowledge Conference}{March 18--22, 2024}{Kyoto, Japan}
\acmBooktitle{The 14th Learning Analytics and Knowledge Conference (LAK '24), March 18--22, 2024, Kyoto, Japan}\acmDOI{10.1145/3636555.3636911}
\acmISBN{979-8-4007-1618-8/24/03}

\begin{document}

\title[Using Think-Aloud Data to Understand SRL Cycles in Intelligent Tutoring Systems]{Using Think-Aloud Data to Understand Relations between Self-Regulation Cycle Characteristics and Student Performance in Intelligent Tutoring Systems}

\author{Conrad Borchers}
\affiliation{
  \institution{Carnegie Mellon University}
  \streetaddress{5000 Forbes Ave}
  \city{Pittsburgh, PA 15213}
  \country{USA}
}
\email{cborcher@cs.cmu.edu}

\author{Jiayi Zhang}
\affiliation{
  \institution{University of Pennsylvania}
  \streetaddress{3101 Walnut St}
  \city{Philadelphia, PA 19104}
  \country{USA}
}
\email{joycez@upenn.edu}

\author{Ryan S. Baker}
\affiliation{
  \institution{University of Pennsylvania}
  \streetaddress{3101 Walnut St}
  \city{Philadelphia, PA 19104}
  \country{USA}
}
\email{ryanshaunbaker@gmail.com}

\author{Vincent Aleven}
\affiliation{
  \institution{Carnegie Mellon University}
  \streetaddress{5000 Forbes Ave}
  \city{Pittsburgh, PA 15213}
  \country{USA}
}
\email{aleven@cs.cmu.edu}

\makeatletter
\let\@authorsaddresses\@empty
\makeatother

\renewcommand{\shortauthors}{Borchers et al.}

\begin{abstract}
Numerous studies demonstrate the importance of self-regulation during learning by problem-solving. Recent work in learning analytics has largely examined students' use of SRL concerning overall learning gains. Limited research has related SRL to in-the-moment performance differences among learners. The present study investigates SRL behaviors in relationship to learners' moment-by-moment performance while working with intelligent tutoring systems for stoichiometry chemistry. We demonstrate the feasibility of labeling SRL behaviors based on AI-generated think-aloud transcripts, identifying the presence or absence of four SRL categories (processing information, planning, enacting, and realizing errors) in each utterance. Using the SRL codes, we conducted regression analyses to examine how the use of SRL in terms of presence, frequency, cyclical characteristics, and recency relate to student performance on subsequent steps in multi-step problems. A model considering students' SRL cycle characteristics outperformed a model only using in-the-moment SRL assessment. In line with theoretical predictions, students' actions during earlier, process-heavy stages of SRL cycles exhibited lower moment-by-moment correctness during problem-solving than later SRL cycle stages. We discuss system re-design opportunities to add SRL support during stages of processing and paths forward for using machine learning to speed research depending on the assessment of SRL based on transcription of think-aloud data.
\end{abstract}



\maketitle

\section{Introduction and background}

Self-regulated learning (SRL) describes the process in which learners take active control of their learning by regulating their attention and effort in pursuit of goals \cite{zimmerman1990self}. A range of cognitive, affective, metacognitive, and motivational processes are involved in SRL, enabling learners to become more independent and effective \cite{zimmerman1990self}. Several models have been proposed to describe SRL from various perspectives \cite{panadero2017review}. Despite their differences, most models describe SRL as a cyclical process consisting of phases where learners understand tasks, make plans, enact the plans, and reflect and adapt. For instance, grounded in information processing theory, Winne and Hadwin \cite{winne1998studying} describe SRL as four interdependent and cyclical stages, in which learners 1) define the task, 2) set goals and form plans, 3) enact the plans, and 4) reflect and adapt when goals are not met. Cognitive and metacognitive SRL strategies, such as monitoring and regulating, are used to achieve the tasks in each stage. Previous studies have shown that successful use of SRL is positively associated with academic achievement \cite{zimmerman2000attaining}. Students who demonstrate more frequent and congruent use of SRL strategies, such as planning and monitoring, are more likely to solve a problem and have better learning outcomes \cite{sabourin2013discovering}.

Given the importance of SRL for learning, past research has leveraged learning analytics to measure and facilitate students' use of SRL in intelligent tutoring systems (ITS), which are personalized, adaptive software for learning, providing step-level guidance during problem-solving \cite{aleven_exampletracing_2016}. Trace data collected from ITS can be used as ``observable indicators to support valid inferences about a learner's metacognitive monitoring and metacognitive control that constitute SRL'' \cite{winne2017handbook}. By analyzing the patterns of behaviors from students interacting with an ITS, researchers can draw inferences, identifying what SRL strategies the students are using, how they are used, and in what order \cite{winne2017handbook, heirweg2020mine, azevedo2010measuring}. These identified SRL behaviors and patterns are often then analyzed in relation to learning and learning outcomes. For example, \cite{segedy2015using} found that students who demonstrated more supported and coherent actions (indicating the use of monitoring) were associated with higher learning gains. Similarly, students who demonstrated more effective help-seeking behaviors (an effective use of SRL) were associated with greater learning gains \cite{aleven2006toward}. This line of research has sparked improved designs of ITS to support the learning of effective metacognitive strategies (e.g., \cite{aleven2006toward}), which in turn has been shown to improve learning by tutored problem-solving, highlighting the importance of SRL in these systems \cite{long2016mastery}. 

Several studies have looked into the cyclical process of SRL to understand how SRL behaviors unfold over time \cite{saint2022temporally}. Methods ranging from statistical models such as hidden Markov models \cite{biswas2010measuring}, to sequence and process mining \cite{sabourin2013discovering, saint2022temporally}, to clustering \cite{zhang2022multi}, have been used to examine the sequential and the temporal order of SRL behaviors. One key hypothesis from this line of work is that the alignment of SRL behavior with theoretical models of SRL cycles (i.e., sequences of processing, planning, and enacting) relate to performance differences during learning \cite{bannert2014process}, tentatively because this congruence relates to how strategically students engage in evaluations of their learning, cognitive, and metacognitive strategies \cite{heirweg2020mine}. Therefore, cyclical SRL models suggest that students who act out goals after engaging in processing and constructing plans are expected to have a higher learning performance compared to students who do not (and act without planning or during processing). Measuring when, during SRL cycles, students act (e.g., by attempting problem-solving steps in ITS) is expected to reveal performance differences, as investigated in the present study.

In addition to log data, think-aloud as another approach to measure SRL in situ has been used in seminal work in SRL \cite{azevedo2008externally}. During think-aloud activities, students are asked to verbalize their thinking and cognitive processes as they interact with an ITS. Utterances collected from think-aloud activities are then coded using a protocol to examine if and how students are engaged in SRL \cite{azevedo2019analyzing}. Thinking aloud neither alters accuracy nor the sequence of operations in most cognitive tasks (excluding insight problems) \cite{fox2011procedures}. Using think-aloud transcripts, Heirweg et al. \cite{heirweg2020mine} show that students commonly adopt a cyclical approach to learning by engaging in preparatory, performance, and appraisal activities. However, high achievers tend to demonstrate more frequent use of SRL, such as planning and monitoring, and are more effective and strategic at implementing SRL strategies \cite{heirweg2020mine}. Additionally, by applying process mining on students' think-aloud, Bannert and colleagues \cite{bannert2014process} show that successful learners engage in SRL behaviors in an order similar to the process theorized by SRL models. In particular, they found that successful learners were likelier to engage in preparatory activities (e.g., orientation and planning) before completing a task. In contrast, preparation and evaluation activities were less frequent in the processing model for less successful students. Similar results were replicated in \cite{lim2021temporal}. 

However, as shown in these studies, SRL has been primarily examined in relation to overall learning gains (e.g., \cite{segedy2015using,aleven2006toward}) or compared between groups (successful vs. less successful learners; e.g., \cite{bannert2014process}). There is a lack of moment-by-moment analyses explicating how the differences in the engagement of SRL relate to the learning and performance at each step during multiple-step problem-solving in an ITS. Few studies are exceptions to this, focusing on in-tutor behaviors such as help-seeking (e.g., \cite{roll2014benefits}). Analyses examining the relationship between SRL and the performance on individual problem-solving steps will provide insight into when and how to provide adaptive interventions that scaffold and facilitate the effective use of SRL during a problem-solving process. Compared to Roll et al. \cite{roll2014benefits}, who analyzed subsequent steps requiring the same skills, we consider correctness along sequences of chronologically ordered problem-solving step attempts during different stages of SRL cycles. Results from this line of research can be used to improve the current SRL interventions that are designed primarily based on the conclusion derived from associating SRL behaviors with overall learning gains (e.g., \cite{lim2023effects}).

To address this gap, this study focuses on relating SRL use with students' step-by-step performance and examining how SRL use (immediately before a step vs. up until a step) is associated with the performance (i.e., correctness) at each step. We collected think-aloud from students working with ITS for stoichiometry chemistry, a domain that, to the best of our knowledge, has not been investigated via fine-grained SRL codes. As suggested in prior work \cite{fan2023towards, rovers2019granularity}, think-aloud protocols provide valid measures on students' use of SRL because they capture an immediate window into students' cognitive and metacognitive processes as opposed to predefined categories in self-report questionnaires. The rich accounts of SRL in think-aloud data are especially useful when limited inferences can be drawn from behavioral logs. Therefore, we generate fine-grained SRL labels for think-aloud utterances related to each step attempt in the ITS. As generating such a high volume of codes is resource-intensive, we transcribe the think-aloud utterances with Whisper \cite{pmlr-v202-radford23a}, a state-of-the-art speech-to-text software, showing that out-of-the-box-transcription can yield sufficiently accurate text data for human labeling of SRL. We combined utterances between two consecutive steps and labeled each combined utterance for four SRL categories, which are 1) processing information, 2) planning, 3) enacting, and 4) realizing errors. These four categories are designed to capture main SRL behaviors within each phase of the four-stage model in \cite{winne1998studying}, representing the cyclical process of SRL. Using the SRL codes, we conducted regression analyses to examine how the use of SRL in terms of presence, frequency, cyclical characteristics, and recency (immediately before a step vs. up until a step) relates to step-by-step performance. To gauge to what extent other linguistic features not represented in our adapted coding scheme might improve inference of problem-solving performance differences and potentially inform revisions of the coding scheme, we further conduct an exploratory analysis of individual words (unigrams) in utterances. Overall, we aim to answer the following three research questions with our analyses:

\begin{itemize}
    \item[] RQ1: What SRL process category (observed immediately before a step) is most related to student performance during learning?

    \item[] RQ2: What features representing SRL cycle characteristics (observed up until a step) relate most to student performance during learning? 
    
     \item[] RQ3: What qualitative insights related to low and high in-tutor performance can be generated from analyzing individual words (unigrams) in utterances beyond SRL process categories?
\end{itemize}

\section{Method}

\subsection{Data collection}

\subsubsection{Study sample}

Ten students enrolled in undergraduate (90\%) and graduate degree (10\%) programs participated in this study between February and May 2023. Five students were recruited at a large private research university (participating in person). The five other students were recruited from a large public liberal arts university (participating remotely via Zoom). All participants were enrolled in degree programs in the United States. Participants were 40\% white, 40\% Asian, and 20\% multi- or biracial. The students were undergraduate sophomores, based on the sample's median year at higher education institutions. Students assessed their prior proficiency in stoichiometry chemistry at an average of 3.4 ($SD$ = 1.35) on a five-point Likert scale ranging from  ``Little prior experience'' to  ``Considerable prior experience''. Students were recruited via course-related information channels via instructors in courses known to include students still learning stoichiometry chemistry and student snowball recruiting in said courses.

\subsubsection{Study materials}

\begin{figure*}[htp]
  \begin{minipage}{\textwidth}
    \centering
    \includegraphics[width=.48\textwidth]{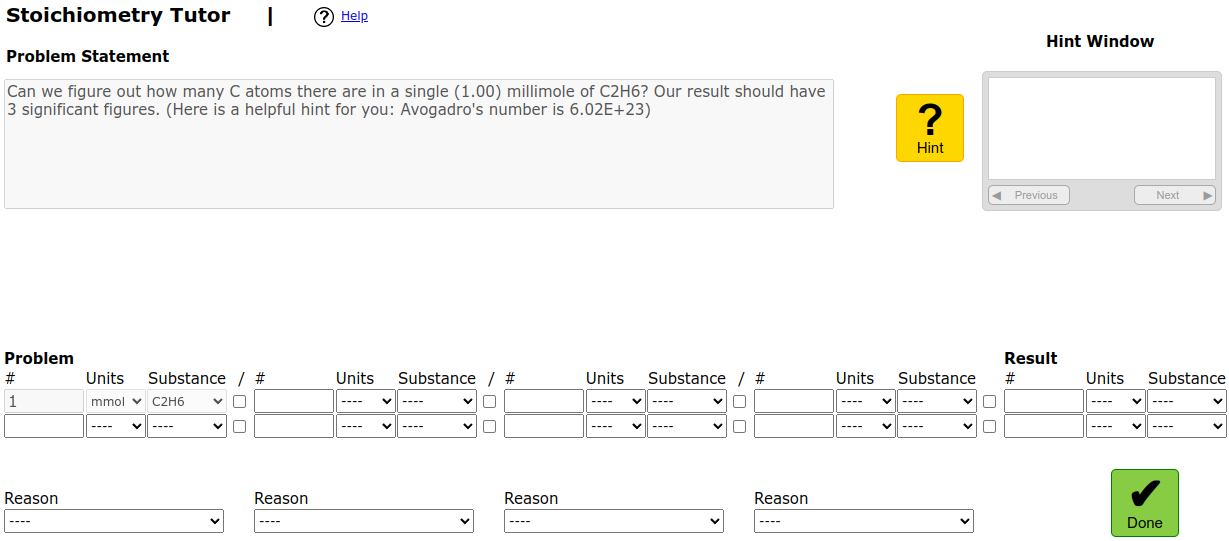}\quad
    \includegraphics[width=.48\textwidth]{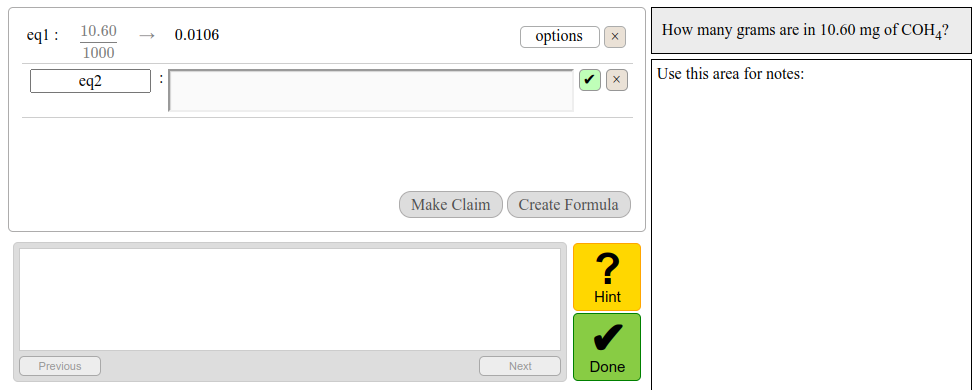}\\
  \end{minipage}
  \caption{Interfaces of Stoichiometry Tutor (left) and ORCCA (right) with two example problem statements.}
  \label{fig:interfaces}
\end{figure*}

Students participating in this study worked with two tutoring systems: the Stoichiometry Tutor and the ORCCA Tutor. Stoichiometry Tutor is a well-studied ITS based on example tracing \cite{aleven_exampletracing_2016}, with evidence that it significantly helps high school students improve in stoichiometry \cite{mclaren2006studying, mclaren2011polite}. Prior work suggests that students in higher education, particularly those not having mastered stoichiometry, likely lack similar content knowledge to that practiced in high school \cite{evans2008learning}. We employed the most recent version of Stoichiometry Tutor, which incorporates findings from prior design experiments, for example, about polite language in the systems's hints \cite{mclaren2011polite}. The interface of the tutoring software is displayed in Figure \ref{fig:interfaces} and characterized by a structured, fraction-based approach to problem-solving, which aims to have the student reason toward a target value. ORCCA Tutor is a novel, rule-based ITS for chemistry \cite{king2022open}. Rule-based ITS match a set of problem-solving rules against a current student's problem-solving sequence to deliver adaptive support based on recognized strategies. Therefore, the ORCCA tutor suits flexible problem-solving sequences with a formula interface (Figure \ref{fig:interfaces}). The ORCCA tutor has not been formally evaluated in peer-reviewed work as of the time of this study. In this work, we analyze data from both tutors jointly based on considerations of statistical power.

A total of eight stoichiometry problems were implemented in both tutoring systems. The problem set originated from prior studies with the Stoichiometry Tutor and was adapted to the rule-based system of the ORCCA tutor. Both tutoring systems and their source code are available upon request.\footnote{\url{https://stoichtutor.cs.cmu.edu/}, \url{https://orcca.ctat.cs.cmu.edu/}}

\subsubsection{Study procedure}

The study lasted 45-60 minutes. It started with a self-paced survey featuring questions about their demographic information, prior experience and grades in academic chemistry, and self-rated proficiency in stoichiometry chemistry. The study included two between-subjects factors. These factors were (a) whether students worked with Stoichiometry Tutor or ORCCA and (b) which content unit students worked on  (i.e., moles and gram conversion vs. stoichiometric conversion). Both content units included a total of four problems. After watching a short introduction video to the relevant tutoring system and being introduced to the think-aloud method, students completed two chemistry problems in their assigned ITS and two more on paper, all while thinking aloud. Then, students completed a post-assessment about their study experience and the usability of the tutor they worked with. If there was sufficient time left (i.e., at least 10 minutes), students would receive an introduction video to the respective other tutoring system and would complete up to two additional problems from the respective other content unit while thinking aloud. Think-aloud utterances were recorded with a 2022 MacBook Pro built-in microphone of the computer serving the tutoring software or Zoom microphones of the participating student's laptop. The ordering of problems was counterbalanced by reversing the problem sequence across all four conditions.

\subsubsection{Analysis data set}

The analysis data set included the log data of individual student transactions (e.g., attempts, hint requests) in the ITS and think-aloud transcripts. Student transactions with each ITS were logged to the PSLC DataShop \cite{koedinger2010data}. Think-aloud transcripts were created using Whisper, a state-of-the-art and open-source transcription model for voice. A comprehensive report on Whisper's accuracy and error types is in \cite{pmlr-v202-radford23a}. Whisper returns segmented utterances with a start and end time stamp. For this study, multiple utterances were combined into one if their timestamp fell between the same two timestamped student actions in the tutoring software (i.e., making attempts and requesting hints). When an action overlaps with an utterance (i.e., an action is carried out while the student is talking), the utterance was joined with the prior utterances. Compared to utterances segmented by Whisper, this concatenation allows coders to examine utterances in a longer time window with more context. Joining utterances based on attempts also allows for modeling the next action's correctness in the tutoring software following SRL codes in the respective prior utterance. 

To model student actions' correctness as a function of SRL codes in think-aloud transcripts, we synchronized the timestamps of both data streams. Reference time stamps were annotated to synchronize data channels. One coder (the first author) familiar with the logging of the tutoring software annotated a reference timestamp for each recording by labeling a transaction in the ITS with a corresponding timestamp in the think-aloud screen capture recording. This process allows for synchronizing tutor log timestamps and recording time stamps with an error of no more than 1 s. Access to the log data and synchronized anonymized think-aloud transcripts is available upon request.\footnote{\url{https://pslcdatashop.web.cmu.edu/DatasetInfo?datasetId=5371}}

\subsection{Coding of SRL components}
We annotated the concatenated utterances with four binary variables representing the presence or absence of four SRL behaviors. For this study, we created a coding scheme adjusted to the tutoring system context based on the four-stage model proposed by Winne and Hadwin \cite{winne1998studying} to reflect the cyclical problem-solving process. Specifically, we coded each concatenated utterance for the following four SRL categories: 1) processing information, 2) planning, 3) enacting, and 4) realizing errors. These four categories reflect a subset of SRL behaviors within each stage of the four-stage model, focusing on the most relevant behaviors within the learning environment being studied (as in \cite{zhang2022detecting, hutt2021investigating}). Similar behavior-centered approaches to coding SRL processes based on the four-stage model have been found useful to predict learning in contexts akin to ITS, that is, learning by problem-solving \cite{hatala2023progression}. As such, the SRL categories in the current study are coarser-grained than other SRL think-aloud studies, which include codes for cognitive and metacognitive operations (e.g., monitoring). However, we opted for a broader coding scheme given that we code think-aloud codes at the level of fine-grained and comparatively short utterances between problem-solving attempts, making cognitive operations that are usually inferred from examining multiple actions and their verbalizations harder to observe. Table \ref{tab:srl} presents the coding categories, including behaviors related to them.

\textit{Processing information} as a critical behavior in the first stage involves learners obtaining and understanding information before they create a mental representation of the task \cite{winne2004students}. During this, behaviors and cognitive activities, such as reading, re-reading the question, and comprehending the question, may happen \cite{winne2004students}. \textit{Planning} captures instances when students set goals and form plans to ensure the progression of problem-solving (e.g., ``I need to do the unit conversion next''). To be included in this category, plans must be conceptual, involve conceptual knowledge, and guide problem-solving. In contrast, procedural planning, such as announcing the next action in the tutoring system (e.g., I'm going to fill out this blank with the unit gram), is coded as \textit{enacting}, as described next. In addition to announcing a concrete action, \textit{enacting} also encodes situations when students verbalize a previous action that has just taken place. Therefore, enacting includes any form of action verbalization with actions corresponding to tutoring software actions (e.g., requesting a hint or entering a specific number). Lastly, \textit{realizing errors} captures instances where students verbalize their thoughts of knowing something is wrong. This realization could be prompted by external factors (e.g., system feedback) or internal reflection \cite{winne2017handbook}. We note that the four SRL categories are not mutually exclusive; that is, a student can demonstrate the use of self-regulation in multiple of the four SRL categories in the same utterance.

\begin{table*}[htp]
\captionof{table}{Overview of the four coded SRL labels, including indicative behaviors of each label and example utterances.}
\label{tab:srl}
\centering
\begin{tabular}{|p{0.12\textwidth}|p{0.545\textwidth}|p{0.265\textwidth}|}
\hline
SRL Category & Behaviors & Example Utterance \\
\hline
Processing\newline information & Assemble information: \begin{itemize}
    \item The utterance demonstrates behaviors where students read or re-read a question, hints, or feedback provided by the system
\end{itemize}
Comprehend information:
\begin{itemize}
\item The utterance demonstrates behaviors where students repeat information provided by the system with a level of synthesis
\end{itemize}
                     & \textit{``Let's figure out how many hydrogen items are in a millimole of water molecule H2O molecules. Our result should have three significant features. Figures. Avogadro number is 6.02 E23. 2 atoms of H2O.''} \\
\hline
Planning &  Identify goals and form plans:
\begin{itemize}
    \item The utterance reflects behaviors where students verbalize a conceptual plan of how they will solve the problem 
\end{itemize}
            & \textit{``Our goal of the result is hydrogen atoms. The goal of the result is the number of hydrogen atoms, right?''} \\
\hline
Enacting & 
Verbalize previous action:
\begin{itemize}
    \item The utterance reflects students' behaviors where they verbalize an action that has just been carried out explaining what they did
\end{itemize}
Announce the next action:
\begin{itemize}
    \item The utterance reflects student behaviors where they verbalize a concrete and specific action that they will do next
\end{itemize}
             & \textit{``Two molecules of this. How many atoms in a... How many atoms in a minimum molecule of M mole? 61023 divided by 2. 3.0115.''} \\
\hline
Realizing errors & 
Realize something is wrong:
\begin{itemize}
    \item The utterance demonstrates instances where students realize there is a mistake in the answer or the process with or without external prompting (i.e., tutor feedback)
\end{itemize}
                     & \textit{``It's incorrect. What's happened? It is the thousand in the wrong spot. 32 grams per mole. No, the thousand is correct, so what am I doing wrong? [...]''} \\
\hline
\end{tabular}
\end{table*}

Using the codebook, the first and second author coded the same 162 concatenated utterances separately to establish reliability. An acceptable inter-rater reliability was established after two rounds of coding ($\kappa_{processing}$ = 0.78, $\kappa_{planning}$ = 0.90, $\kappa_{enacting}$ = 0.77, $\kappa_{error}$ = 1.00). The coders then split the rest of the utterances and completed the coding individually. Utterances from previous coding iterations with no sufficient reliability were double-coded.

\subsection{Data analysis methods}

\subsubsection{SRL variables and performance outcome variable}
\label{sec:feateng}

Research questions 1 and 2 relate to the associations of in-the-moment and historical features of student SRL processes with student performance during problem-solving. In both cases, the modeling task is to model the correctness of student attempts in the tutoring system by their preceding SRL codes based on student think-aloud utterances. Related to RQ1 and RQ2, we engineer two broad classes of features from our SRL codes, which we separate into two categories: in-the-moment SRL features and SRL cycle characteristics features. In-the-moment SRL features include characteristics of utterances right before a given attempt in the tutoring system. In particular, the features include four binary features representing the presence or absence of each of the four SRL codes. SRL cycle features represent characteristics of the student's SRL process until the point of the subsequent attempt. This allows for encoding SRL sequence loop characteristics, which prior work found to relate to problem-solving performance and proficiency \cite{hatala2023progression,lim2021temporal}. 

For SRL cycle characteristics features, we create a categorical variable representing the students' current SRL loop state before an attempt in the ITS. An SRL loop is initialized by an utterance with the ``process'' label and closed when an initialized loop has been followed by both a ``plan'' and an ``act'' label. Loops are only closed once planning, and then acting, have occurred after the process label; in other words, ``process, act, plan, act'' would count as a loop, but not ``process, act, act, plan''. All other attempts are considered outside of SRL loops; all utterances with SRL sequences that do not return to processing after closing a loop via acting, such as ``act, plan, act'', are not considered loops. The SRL loop state variable has the levels ``inside loop'' and ``out of loop''. This feature is motivated by \cite{lim2021temporal} showing that successful students showed more congruence to sequences of preparing, planning, and enacting (i.e., ongoing loops in our study), contrasting to less successful learners who demonstrated difficulty in combining these SRL behaviors, leading to detached SRL patterns (i.e., out-of-loop codes in our study). Similarly, features representing the length of SRL loops may distinguish between levels of in-tutor performance during loops. One recent study \cite{hatala2023progression} detected groups of SRL sequences in higher education programming assignments. The resulting SRL sequences had varying levels of iterations of the center components of SRL sequences (i.e., longer loops) and significantly distinguished between proficiency levels. Based on this finding, we engineer two features representing each student's SRL loop characteristics. First, we computed the student's current number of attempts since the last closed loop (``unclosed since''), representing the ongoing loop's length. Second, we computed the number of attempts per loop, representing the average length of students' SRL cycles updated at each attempt. Note that both variables are computed only during ongoing loops and set to 0 outside of loops to capture differences in students' SRL loops during modeling of student in-tutor performance.

As our primary outcome variable for the study, we analyze students' correctness at attempts, with hint requests considered as evidence of not knowing a particular skill and coded as an attempt, in line with standard practice for analyzing log data from ITS
\cite{chi2011ifa, koedinger2010data}.

\subsubsection{Model comparisons and statistical tests}

Our first two research questions, RQ1 and RQ2, relate to the general inferential utility of our coded SRL categories concerning in-tutor correctness. We opted for inferential regression over predictive out-of-sample modeling approaches to focus on deriving insight into associations and the relative strength of relationships between SRL features and in-tutor performance. Compared to complex machine learning architectures, regression does not pose a risk of overfitting to our sample and allows for the most straightforward interpretability of our results (as methods to achieve explainability for machine learning architectures remain contradictory and complicated \cite{swamy2022evaluating}). We contrast two sets of features for each RQ: in-the-moment SRL assessment right before problem-solving attempts in the tutor and SRL cycle characteristics (see Section \ref{sec:feateng}) via likelihood-ratio tests, as described next.

We compare three logistic regression models: A null model featuring simply a random intercept for the baseline correctness of students against a model featuring the four binary variables of given utterances representing the presence or absence of the four SRL codes. This model is then compared to a model additionally featuring our three historic SRL variables: the cycle state, the number of unclosed attempts since utterances, and the current average number of attempts per loop. Given that the correctness of attempts is not independent within students, which may bias $p$-values of model coefficients toward significance, we additionally add a random student intercept to each model, adjusting for each student's baseline correctness in the tutor via generalized linear mixed modeling \cite{bolker2015linear}. All independent variables were standardized to a mean of 0 and a standard deviation of 1 to aid model coefficient interpretation. We report odds ratios ($ORs$) instead of logarithmic $\beta$ coefficients in logistic regression for all estimated effect sizes to further aid interpretation.

$ORs$ represents the factor by which two odds ($\frac{p}{1-p}$) of getting an attempt in the tutoring software right is smaller or larger per standard deviation (numeric features) or presence (categorical features) of a variable. For example, an $OR$ of 2 indicates that the odds of getting an attempt correct are twice as large if the ``planning'' code was present in the previous utterance, while an $OR$ of 0.5 indicates that it would be half as large. We note that odds ratios can not be interpreted as frequencies, that is, correct attempts occurring twice or half as often. To rule out multicollinearity, we further ensured that all model features have variance inflation factors below 2 \cite{craney2002model}.

\subsubsection{Exploratory and qualitative data analysis}

Combined utterances may encode more linguistic information relevant to inferring in-tutor correctness than our assumed SRL categories. Therefore, we mine unigram features that may reliably distinguish between utterances before incorrect and correct attempts to investigate RQ3. We employ $\chi^2$-based unigram sampling, which filters the most distinguishing words between two categories of documents (i.e., the correctness of the subsequent attempt). Addressing potential spelling differences in Whisper's word-level accuracy \cite{pmlr-v202-radford23a}, we apply standard preprocessing techniques during tokenization (i.e., making words lowercase and removing punctuation). The resulting number of tested unigram pairs is $N$ = 838. Therefore, we adjust the Type I error level $\alpha$ via the Benjamini-Hochberg procedure at a false-discovery rate of 0.05 \cite{benjamini1995controlling}. All data analysis code is publicly available.\footnote{\url{https://github.com/conradborchers/srl-cycles-lak24}} 

\section{Results}

\subsection{RQ1: In-the-moment associations of SRL codes with in-tutor correctness}

RQ1 pertained to the associations of the presence of different SRL codes with the correctness of subsequent attempts in the tutoring system. Out of 630 attempts in the tutoring system, 96 (15.24\%) attempts were immediately preceded by processing, 70 (11.11\%) by planning, 137 (21.75\%) by acting, and 37 (5.87\%) by the wrong code. Finally, 360 (57.14\%) attempts had none of the SRL codes in their respective preceding utterances. A model featuring binary variables representing the presence and absence of each code significantly fit the data better than the null model with a simple intercept based on a likelihood-ratio test, $\chi^2(4) = 16.17, p = .003$. An overview of the selected model is in Table \ref{tab:m1}. 

\begin{table}[htp]
\caption{Linear mixed logistic regression with an individualized student intercept using in-the-moment features of SRL from think-aloud utterances to infer next-attempt correctness with odds ratios ($OR$) larger than 1 representing higher correctness ($N=630$).}
\label{tab:m1}
\begin{tabular}[t]{lccc}
\toprule
Effect & $OR_{correct}$ & $CI_{95\%}$ & $p$\\
\midrule
Intercept                 & 0.71 & [0.41, .123] & .225\\
Processing  & 0.53 & [0.31, 0.91] & .014\\
Planning         & 1.05 & [0.59, 1.88] & .869\\
Enacting    & 1.55 & [1.01, 2.37] & .046\\
Realizing errors       & 0.34 & [0.14, 0.84] & .015\\
\bottomrule
\end{tabular}
\end{table}

\begin{figure*}[htp]
    \centering
    \includegraphics[width=0.8\textwidth]{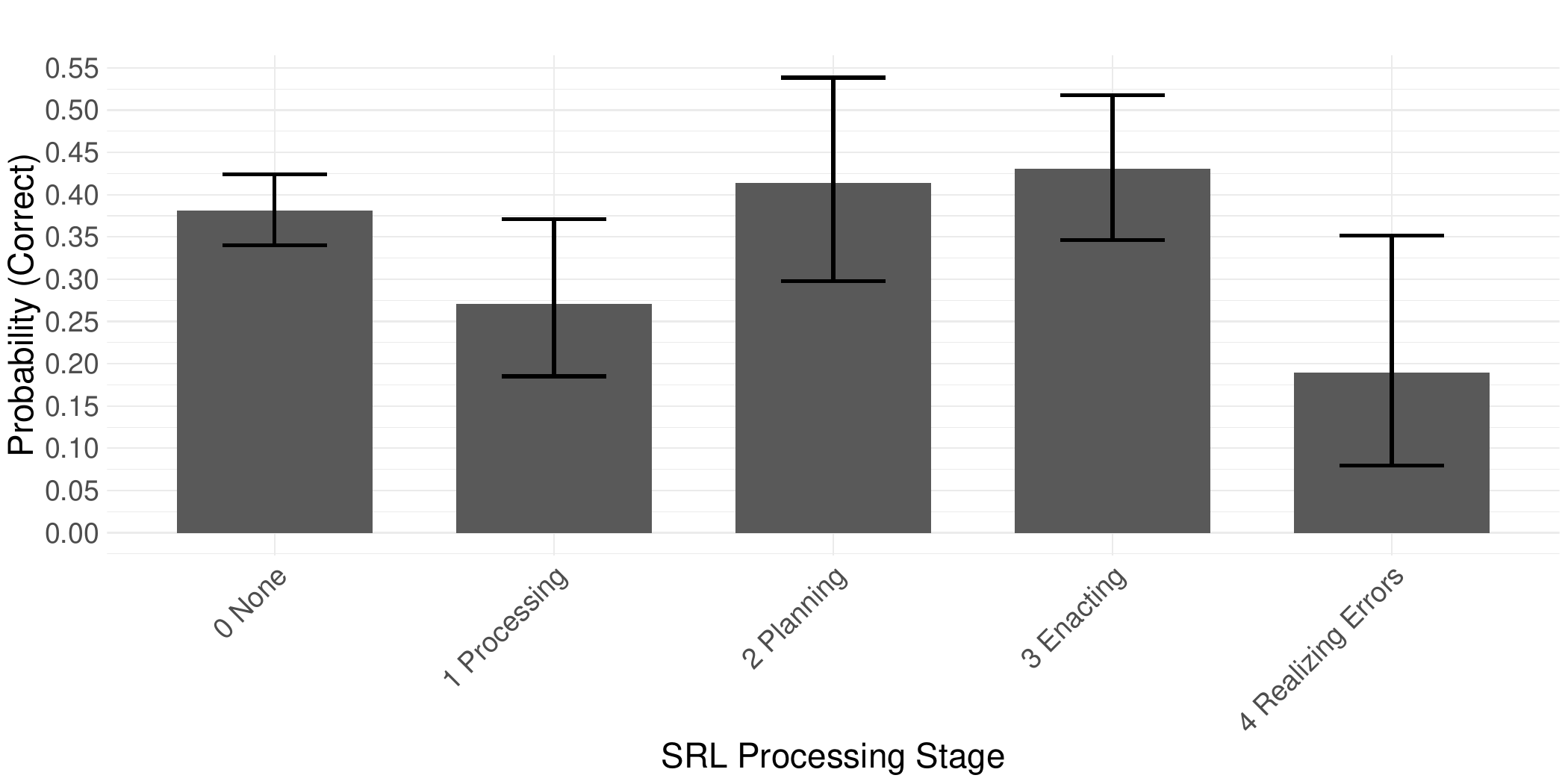}
    \caption{In-tutor correctness at subsequent attempts by the presence of different SRL stages, including a control group of utterances without any SRL codes and 95\% binomial confidence intervals.}
    \label{fig:binomial}
\end{figure*}

As shown in Table \ref{tab:m1}, students had a significantly lower probability of having a correct attempt in the system after utterances with codes for processing and realizing errors in them. Conversely, students had a significantly higher probability of having a correct attempt in the system after utterances with enacting in them. The effect of planning on correctness was not significant ($p = .869$). Univariate associations of each code based on $\chi^2$ tests of independence with correctness aligned (with the exception of processing and enacting, which were only marginally significant instead of significant) with these results for processing, $\chi^2(1) = 3.82, p = .051$, planning, $\chi^2(1) = 0.62, p = .432$, enacting, $\chi^2(1) = 2.95, p = .086$, and realizing errors, $\chi^2(1) = 4.44, p = .035$. Overall, these differences reflected that students' in-tutor correctness was lowest following the presence of SRL codes related to stages of processing instruction and tutor feedback, that is, processing and realizing errors (Figure \ref{fig:binomial}).

\subsection{RQ2: Inferential utility of historical beyond in-the-moment SRL features}

To investigate RQ2, we constructed a model featuring SRL cycle characteristics in addition to in-the-moment SRL features used in the previous model. The model with SRL cycle characteristics significantly fit the data better than the model using only in-the-moment SRL codes right before attempts based on a likelihood-ratio test, $\chi^2(3) = 14.08, p = .003$. An overview of the selected model is in Table \ref{tab:m2}.  

\begin{table}[htp]
\caption{Linear mixed logistic regression with a random student intercept using in-the-moment and historical features of SRL from think-aloud utterances to infer next-attempt correctness with odds ratios ($OR$) larger than 1 representing higher correctness ($N=630$).}
\label{tab:m2}
\begin{tabular}[t]{llcc}
\toprule
Effect & $OR_{correct}$ & $CI_{95\%}$ & $p$\\
\midrule
Intercept                 & 0.83 & [0.46, 1.50] & <.001\\
Processing  & 0.54 & [0.31, 0.96] & .035\\
Enacting    & 1.56 & [1.01, 2.41] & .045\\
Planning         & 0.98 & [0.54, 1.77] & .951\\
Realizing errors       & 0.30 & [0.12, 0.76] & .011\\
\textit{Loop State [vs. Out-the-Loop]}\\
\hspace{2mm}In-the-Loop  & 0.79 & [0.46, .135]  & .395 \\
N Unclosed Since        & 1.33 & [1.01, 1.73] & .039\\
Attempts per Cycle  & 1.48 & [1.15, 1.89] & .002\\
\bottomrule
\end{tabular}
\end{table}

Interpreting coefficients in Table \ref{tab:m2}, there were two notable associations between SRL characteristics and correctness. First, more attempts per cycle (i.e., longer cycles) were associated with higher correctness on subsequent attempts ($OR$ = 1.48, $p$ = .002). Second, correctness was significantly higher later during students' SRL cycles, that is, the longer the start of an ongoing cycle was away in terms of number of attempts ($OR$ = 1.48, $p$ = .002). All other SRL cycle features were not significant. Student intercepts explained $R^2 = 14.0\%$ of the variance in correctness beyond fixed effects ($R^2 = 6.5\%$).

\subsection{RQ3: Qualitative insights from further linguistic analyses}

To investigate RQ3, we mined additional features that related to in-tutor correctness. We present statistical tests of distinct words being able to distinguish utterances by whether subsequent attempts were correct or incorrect (Table \ref{tab:m3}).

\begin{table*}[htp]
\caption{Top ten most distinct unigrams associated with correct (left) and incorrect subsequent attempts (right). $N_{correct}$ and $N_{incorrect}$ denote how frequently each unigram occurred in utterances preceding correct and incorrect attempts, respectively.}
\label{tab:m3}
\begin{tabular}{rrrrr|rrrrr}
Unigram & \textbf{$N_{correct}$} & \textbf{$N_{incorrect}$} & \textbf{$\chi^2$} & \textbf{$p$} & Unigram & \textbf{$N_{correct}$} & \textbf{$N_{incorrect}$} & \textbf{$\chi^2$} & \textbf{$p$} \\
\hline
p4            & 7                    & 0                      & 12.55      & <.001   & hydrogen      & 8                    & 51                     & 12.70      & <.001   \\
good          & 11                   & 4                      & 9.19       & .002   & liters        & 1                    & 21                     & 9.35       & .002   \\
every         & 5                    & 0                      & 8.96       & .003   & point         & 4                    & 29                     & 8.05       & .005   \\
denominator   & 6                    & 1                      & 7.59       & .006   & glucose       & 8                    & 37                     & 6.36       & .012   \\
co2           & 4                    & 0                      & 7.17       & .007   & whats         & 0                    & 11                     & 6.14      & .013   \\
fe2o3         & 4                    & 0                      & 7.17       & .007   & kilo          & 3                    & 20                     & 5.18       & .023   \\
right         & 22                   & 19                     & 5.69       & .017   & entry         & 0                    & 6                      & 3.35       & .067   \\
far           & 3                    & 0                      & 5.38       & .020   & produce       & 1                    & 9                     & 2.90       & .089   \\
somewhere     & 3                    & 0                      & 5.38       & .020   & per           & 31                   & 37                     & 2.83       & .092   \\
thousandth    & 3                    & 0                      & 5.38       & .020   & around        & 0                    & 5                      & 2.79       & .095  
\end{tabular}
\end{table*}

In line with Table \ref{tab:m3}, abstract formula representations of substances (e.g., p4, co2) were more indicative of subsequent correct attempts, while commonplace signifiers of substances (e.g., hydrogen, glucose) were more indicative of subsequent incorrect attempts. However, after adjusting the Type I error level $\alpha$ via the Benjamini-Hochberg procedure at a false-discovery rate of 0.05, none of the tested pairwise comparisons were statistically significant.

\section{Discussion}

The present study investigated the temporal associations of self-regulated learning (SRL) codes based on automated think-aloud transcriptions with problem-solving performance in intelligent tutoring systems (ITS) for stoichiometry chemistry. Our results contribute to the scientific understanding of fine-grained associations of SRL processes with problem-solving performance during tutoring. Furthermore, our findings demonstrate that unaltered, automated transcriptions of think-aloud recorded with laptops during problem-solving have sufficient fidelity to be reliably coded for SRL processes by humans. Finally, our comparisons of in-the-moment and historical features guide the development of future learning analytics to support SRL processes based on live SRL assessment via think-aloud methodologies.

Our first finding relates to in-the-moment differences in problem-solving performance as a function of SRL codes (RQ1). We found in-tutor correctness to be significantly lower on attempts after utterances with codes of processing and realizing errors and significantly higher on attempts after utterances, including enacting. All associations align with a cyclical SRL process model that predicts performance differences between students when acting follows stages of processing and planning, as opposed to during processing \cite{bannert2014process}. Why did students who acted during the processing stages of SRL exhibit lower correctness? Past work suggests one potential reason: these students might have been overall less strategic in their approach to problem-solving, not engaging in evaluations of their cognitive and metacognitive strategies \cite{heirweg2020mine}. Consequently, they would generally exhibit less processing and planning before eventually acting out attempts in the ITS. If students reflect less about their problem-solving attempts, we would expect these students to generally show less learning (in terms of knowledge gains and learning rates in the system, that is, improving over time). However, our current sample does not allow us to further test this hypothesis, which poses fruitful future work, as similar effects of knowledge gains have been documented outside of ITS \cite{segedy2015using,lim2021temporal}. Alternatively, it could be that student disengagement or lack of motivation, although less likely during supervised think-aloud sessions due to the presence of an experimental conductor, could explain why some students acted more (and had lower performance in the ITS) rather than processing and planning ahead of time. Future work could employ affect detectors \cite{pardos2013affective} to further investigate this hypothesis. A third explanation could relate to students having difficulties operating the tutoring system's interface, therefore not being able to act out plans created during SRL processes properly (or trying out the interface to learn it, leading to more errors), therefore acting more prematurely according to log data. Prior work has documented related issues in interface learning and transfer to new learning environments after ITS practice (e.g., \cite{borchers2023makes}), and future work could further investigate this explanation by sampling learners over a longer time frame than in the present study, which spanned a single session of less than 60 minutes per learner. Overall, our results extend prior work showing that students with cyclical SRL behavior show improved learning outcomes \cite{bannert2014process, lim2021temporal} by offering first-of-its-kind evidence that SRL cycle characteristics are associated with moment-by-moment learner performance differences in intelligent tutors.

Our results also add to the scientific understanding of prior work on learning in ITS. Students having lower correctness after realizing errors can be interpreted as students continuing to make errors after prior errors. At least one prior study found that errors in tutoring software tend to be followed by further, subsequent errors \cite{aleven2000limitations}. Potential reasons cited include a lack of help-seeking by the student after errors. Therefore, one implication of our results is to improve the current tutoring software design by providing additional scaffolds for SRL after errors, which guide the student to process instruction and error feedback rather than continuing to act. Future work could take inspiration from past studies showing learning benefits from adaptive SRL scaffolds (e.g., \cite{roll2011improving,lim2023effects}. Such approaches to re-design could complement past frameworks of data-driven re-design for ITS that leverage log data based on cognitive modeling \cite{huang2021general}.  

Bolstering the interpretation that considering SRL cycles helps in predicting in-tutor correctness, our results suggest additional features representing SRL cycle characteristics describe moment-by-moment performance differences better than in-the-moment SRL assessments alone (RQ2). Specifically, we found the number of attempts since the last SRL loop to significantly positively relate to in-tutor correctness. Assuming a cyclical model of SRL where students follow sequences of processing, planning, and acting, the number of attempts since the last started SRL loop represents how far away students are from the initial processing stages related to the given step or how close (in terms of attempts) they are to completing a cycle via verbalizing an action. Therefore, a positive association between the attempts since the last cycle started and correctness suggests that students made more errors during the SRL cycle's initial (presumably more process-heavy) stages. This aligns with the RQ1 finding highlighting how processing was generally negatively related to correctness. Yet, the finding also suggests that assessing a student's SRL cycle in addition to the in-the-moment presence of processing offers a more accurate understanding of learner performance. Therefore, one design implication for future work is that adaptive scaffolds for SRL should not only jump in when a stage of processing is detected but should explicitly consider students' SRL cycle characteristics. Such scaffolds could also signal to the student what the next stage in their SRL cycle should be, for example, encouraging planning instead of acting after processing if a student tends not to plan. Such scaffolds could draw inspiration from past work using scaffolding for planning that successfully supported learners working with ITS \cite{azevedo2022lessons}. One example adaptation could be to display the present subgoal of problem-solving in a textbox in the interface to remind learners who tend to act during processing about existing plans. Alternatively, learners could occasionally be prompted to set new subgoals after correct attempts. As stated in previous literature \cite{lim2023effects}, one key aspect in the design of such SRL interventions is its ability to fade scaffolding out. Adaptivity and prompts suggested by the system could aim to provide initial scaffolding to guide and promote students' ability to regulate and increasingly fade out as students gain the ability to effectively self-regulate their learning.

We also found that the number of attempts per cycle (i.e., the average length of cycles at a given step) positively relates to correctness. In other words, students with longer SRL cycles, as measured in how many attempts students have before closing cycles via action verbalization, had higher correctness in the tutor. This finding is incompatible with a theoretical model where students with coherent and uninterrupted (and therefore short according to our definition) sequences of processing, planning, and acting exhibit better performance \cite{lim2021temporal, li2022temporal}. However, as the estimate of cycle length gets updated at every step, cycle length tends to be higher during later rather than earlier stages of SRL cycles. Therefore, the effect of cycle length might also be partially explained by earlier stages of processing being negatively related to performance. More work is needed to replicate this effect and see if this pattern is seen consistently in other data. Alternatively, it could be the case that students with higher performance verbalize acting less due to higher task fluency. Consequently, their think-aloud utterances would include fewer instances of action verbalizations, which are required to close SRL loops, prolonging their assessed average loop length. While our linear mixed models generally adjust for baseline differences in in-tutor correctness via individualized student intercepts, we can not rule out this possibility because the present sample size is insufficient to estimate individualized slopes of the loop length effect per student, which future work could correlate with baseline correctness to test this explanation. Future work could also conduct a more in-depth analysis of how the acting category can be assessed. One potential direction could be to assign an acting code when a plan expressed during planning stages is carried out in the tutoring software, for example, by relating a plan to a specific input field in the software interface. Alternatively, future work could experiment with less conservative definitions of SRL cycles (where repeated instances of planning without acting or acting without planning are not strictly considered to be ongoing loops, as we do in the present study).

Rich linguistic features in student think-aloud processes between attempts might bear more insight into learning than SRL codes. Therefore, our third research question concerned qualitative insights into single words (unigrams) in utterances related to in-tutor correctness. However, our analysis suggested that none of the unigram features we tested significantly related to next-attempt correctness after adjusting for multiple tests. As several of the extracted unigrams were domain-specific (e.g., representing substances such as co2) while our SRL coding scheme is domain-agnostic, we hypothesize that variation in student performance related to SRL can be sufficiently captured via domain-agnostic SRL coding schemes for think-aloud utterances. However, more research is needed to replicate our procedure and investigate if more sophisticated features based on natural language other than unigrams (e.g., n-grams, sentence embeddings, or sentiment analysis) could shed more light on linguistic associations of think-aloud data with in-tutor correctness. Related work on learning in MOOCs showed some merit in predicting learner performance from such linguistic features extracted from discussion posts \cite{duru2021deep}. Another area of future work is to investigate if the use (or lack) of domain-specific vocabulary can guide adaptivity in tutoring software. Such diagnostics could improve tutor support (if vocabulary is part of the system's learning objective) or re-design the language in problem statements to be more individualized, which prior work has found could improve student learning outcomes \cite{walkington2013supporting}. Future work could leverage both think-aloud data and other text artifacts, such as self-explanations, often part of tutoring software's learning objectives, to expand on this research direction.

\subsection{Limitations and future work}

We see four limitations to our present study that guide future work. First, the current sample is limited in estimating the generalizability of our SRL models to different learner tasks and contexts. The present study investigated two ITS for high school and remedial college stoichiometry chemistry, which is a domain that, to the best of our knowledge, has not been investigated in terms of moment-by-moment associations between SRL and in-tutor performance. Yet, one of the systems, the ORCCA tutor, allows for more open-ended responses and scaffolds the problem-solving process less than Stoichiometry Tutor. Less problem-solving scaffolding means that learners may have to rely more on SRL processes to solve a problem, such that SRL processes are expected to have more weight on performance differences across students \cite{winne2010improving}. Our current sample size does not allow us to estimate tutor-specific models of SRL reliably to investigate the effects of these differences in scaffolding on learning, which should be an area of future work. Similarly, it is an open question to what extent models of SRL based on this study will generalize to domains outside of chemistry, for example, ITS for mathematics. Future work may investigate how studying an SRL model on different tutoring software within and across domains generalizes in its ability to infer problem-solving performance, informing model training needs for adapting our methodology to new systems.

Second, the present study's coding scheme and mode of tagging utterances with SRL codes calls for further investigation. Specifically, future work is needed to gauge to what extent the present study's SRL labels can be reliably predicted in text classification tasks, for example, by using BERT, which has been fruitfully used for similar prediction tasks in learning analytics (e.g., \cite{goslen2023enhancing}), or recent large language models. Future work could also investigate if the SRL codes used in this study can be reliably predicted from automated transcriptions of think-aloud, which we used to automate the process of transcribing think-aloud for human labeling, which was still labor-intensive. Research could also explore pre-processing techniques for automated transcripts to improve prediction, as automated transcription itself is imperfect and introduces noise to the text data that serves as the basis of prediction. Future work may also replicate our methodology to test if out-of-the-box transcription of languages other than English is sufficiently accurate to allow coders to reliably code SRL categories. As of the time of writing this manuscript, Whisper supports 99 languages.

Third, more work is needed to investigate how fine-grained codes of SRL during learning by problem-solving can be used to improve live adaptivity and student-facing analytics in ITS. The present study suggested some pathways for tutor re-design, including additional SRL scaffolds during the initial processing stages. Yet, even if processing could be reliably detected from live transcriptions of think-aloud data and machine learning, it is an open question to what extent such adaptivity pipelines are practical. Having multiple students concurrently think aloud in the same room in larger classrooms is not feasible. However, there might be other opportunities to collect think-aloud data during classroom instruction, for example, during teacher visits in the context of individualized problem-solving, where teachers ask students to walk them through a problem or thought process. Another potential remedy is having students complete a pre-assessment or parts of tutor practice at home, where audio data can be collected without disturbing other learners. Future work is encouraged to investigate whether learners can be automatically classified into learner profiles via live think-aloud transcription and machine learning, which could then guide SRL scaffolds in subsequent practice units. Intermediate assessments could then be used to ascertain whether and when these scaffolds can be faded, a strategy that prior work identified to benefit learners \cite{puntambekar2005tools}. Such adaptive policies could be compared to log data-based models of SRL prediction to guide SRL scaffolds and non-adaptive fading policies for SRL scaffolds \cite{azevedo2004does, azevedo2022lessons, munshi2023analysing}. Finally, we note that trace data models predicting SRL labels from our sample might achieve live adaptivity that improves learning without concurrent think-aloud. Similar work demonstrates the viability of such approaches in eye-tracking \cite{conati2020comparing}.

Fourth, the SRL categories identified in the current study primarily stem from think-aloud data, representing a singular data stream for assessing students' SRL. Think-aloud has been commonly used in previous literature to evaluate SRL. However, more recent studies have investigated and demonstrated the possibility of measuring SRL through multimodal data, incorporating sources such as behavioral logs, eye tracking, and physiological data such as heart rate and skin galvanic conductivity \cite{molenaar2023measuring,azevedo2019analyzing}. The process of SRL might be more comprehensively captured by employing diverse measurement approaches. In our future research, we intend to explore the potential of integrating log data into SRL measurement to understand how SRL cycles are associated with in-tutor performance.

\section{Summary and Implications}

The present study demonstrated the feasibility of hand-coding reliable SRL codes from off-the-shelf think-aloud transcriptions that relate to student performance during problem-solving practice. Our findings advance the scientific understanding of how cyclical SRL features explain performance differences during problem-solving in intelligent tutoring systems. We find that students exhibited higher levels of correctness on problem-solving attempts following action verbalization, as opposed to during more initial stages of processing and after realizing errors. Specifically, the farther along in ongoing SRL cycles students were, the more likely they were to get subsequent steps in the tutoring system correct. These empirical findings contribute evidence for a cyclical model of SRL, which also yielded a significantly more accurate modeling of the effects of SRL on learner performance. However, contrary to recent research, our findings do not suggest that students with shorter, uninterrupted SRL cycles exhibited higher correctness. More work is needed to investigate this finding, potentially by extending our current operationalization of closing SRL loops via action verbalization. For practice and system design, our results imply that intervening when students tend to attempt problem-solving steps during more initial stages of processing poses an opportunity for tutoring system re-design and improved SRL scaffolds. Predicting SRL processes from live transcriptions via machine learning may be a relevant step towards creating analytics-based interventions for students that complement other approaches to tutoring system re-design.

\begin{acks}
Carnegie Mellon University's GSA/Provost GuSH Grant funding was used to support this research. Funding to attend this conference was provided by the CMU GSA/Provost Conference Funding.
\end{acks}

\bibliographystyle{ACM-Reference-Format}
\bibliography{main}

\end{document}